\documentclass[12pt]{article}
\usepackage{amsmath}
\usepackage{graphicx,psfrag,epsf,fullpage}
\usepackage{enumerate}
\usepackage{natbib}
\usepackage{url} 
\usepackage{multirow, booktabs}
\usepackage{amsthm,amssymb,bm}
\usepackage{setspace}
\usepackage[dvipsnames]{xcolor}
\allowdisplaybreaks[4]

\newtheorem{thm}{\bf Theorem}[section]
\newtheorem{assumption}{Assumption}[section]
\usepackage{hyperref} 
\usepackage{cleveref} 
\numberwithin{equation}{section}

\hypersetup{colorlinks = true, 
	    linkcolor = black, 
	    urlcolor = blue,
            citecolor = black} 
\usepackage{subfig,graphicx}

\begin{document}

\def\spacingset#1{\renewcommand{\baselinestretch}%
{#1}\small\normalsize} \spacingset{1}


  \title{\bf  Adaptive Rank-based Tests for High Dimensional Mean Problems}
  \author{Yu Zhang and Long Feng\thanks{Corresponding author: flnankai@nankai.edu.cn}\\
   School of Statistics and Data Science, KLMDASR, LEBPS, and LPMC,\\
   Nankai University, No.94 Weijin Road, 300071, Tianjin, China}
  \maketitle

\begin{abstract}
The Wilcoxon signed-rank test and the Wilcoxon-Mann-Whitney test are commonly employed in one sample and two sample mean tests for one-dimensional hypothesis problems. For high-dimensional mean test problems, we calculate the asymptotic distribution of the maximum of rank statistics for each variable and suggest a max-type test. This max-type test is then merged with a sum-type test, based on their asymptotic independence offered by stationary and strong mixing assumptions. Our numerical studies reveal that this combined test demonstrates robustness and superiority over other methods, especially for heavy-tailed distributions.
 \end{abstract}

\noindent%
{\it Keywords:}  Asymptotic independence, High dimensional test, Rank-based test.
\vfill

\newpage

\section{Introduction}
\label{sec:intro}
In the past few decades, hypothesis test for mean has drawn a significant amount of attention in a wide range of scientific fields, such as chemistry \citep{miller1991basic}, education \citep{cheung2009students}, genetics \citep{ouyang2022rank}, economics \citep{newbold2013statistics} and many other fields. Hotelling’s $T^2$ test is the classic method for mean test problem, which does not work in high dimensional problem because the sample covariance matrix is not invertible. A large amount of efforts have been made for replacing the sample covariance matrix in the classic Hotelling's $T^2$ test statistic. For example, \cite{bai1996effect}, \cite{chen2010two} replaced the singular matrix with identity matrix in Hotelling’s $T^2$ test. \cite{srivastava2008test} do the replacement with the diagonal matrix of sample covariance matrix. More  replacement-based work can see the \cite{feng2015two}, \cite{feng2016multivariate}, \cite{gregory2015two}, \cite{park2013test} and \cite{wang2015high}. The above methods can be viewed as sum-type methods, which consist of summation of parameter estimators. The mentioned analytics show that the sum-type tests generally perform well as data are dense, where dense means most of the parameters are nonzero under the local alternative. However, \cite{tony2014two} mentioned that sum-type statistics does not perform as well in sparse cases as in dense cases, where sparse means only a few of the parameters are zero under the local alternative. To handle the sparse alternatives, \cite{chen2019two}, \cite{feng2022asymptotic} and \cite{tony2014two} proposed some max-type tests, which performs well under sparse alternatives.

In order to adopt appropriate methods for high dimensional hypothesis mean test, it's essential to get a rough picture of alternatives. However, we can not know whether the data are sparse or dense. Recently, many literatures proposed adaptive methods to overcome this issue, such as \citet{fan2015power}, \citet{xu2016adaptive}, \citet{he2021asymptotically}, \cite{feng2022asymptotic}. For example, \cite{feng2022asymptotic} proposed a data-robust test which combines the sum-type test and max-type test under Gaussian assumptions, performs pretty well under both sparse and dense alternatives. It proves the asymptotic independence between the sum-type statistic and max-type statistic, which motivates us to construct combination test based on independence between the different type tests.

Indeed, the aforementioned methods are all extensions of the classic $t-$test for a one-dimensional variable. It’s well-known that the classic $t-$test does not perform very well for heavy-tailed distributions. Therefore, the Wilcoxon signed-rank test and the Wilcoxon-Mann-Whitney test are typically used in one sample and two sample mean tests for one-dimensional hypothesis problems.
\cite{ouyang2022rank} expanded these two rank-based methods to high-dimensional mean problems by summing the test statistics of each variable, which has proven effective for dense alternatives. For sparse alternatives, we introduced a new max-type test based on rank statistics and established its asymptotic distribution.
Subsequently, we combined this newly proposed max-type test with the sum-type test proposed by \cite{ouyang2022rank}, under the $\alpha$-fixing and stationary conditions provided by \cite{hsing1995note}, for both one sample mean test and two sample mean test.
Our simulation results indicate that our proposed combination tests not only inherit the robustness of signal sparsity, but also for the heavy-tailed distributions.

The rest of the paper is organized as follows. In section 2, we review the moment properties of  Wilcoxon signed-rank test, Wilcoxon Mann-Whitney test and state the theoretical results of our proposed max-type test and combination test for one sample case and two sample case respectively. Simulation studies are shown in Section 3. In section 4, we make conclusions about this paper. Finally we present the technical proof in the Appendix.

\section{High dimensional mean test}
\label{sec:meth}

\subsection{One sample test}
Assuming $p$ dimensional vectors $\bm{X}_1,...,\bm{X}_n$ generated independently from distribution $F(\bm{X}-\bm{\mu})$, where the  distribution of $F$ is continuous and symmetric. For the one sample test problem, the hypothesis problem can be expressed as:
\begin{equation}\label{eq1}
    H_0:\bm{\mu}=0  \ \   versus\ \ H_1:\bm{\mu}\neq 0.
\end{equation}
In the case of $p=1$, it's natural to consider the Wilcoxon signed-rank test. The statistic of Wilcoxon signed-rank is defined as:
\begin{equation}
    U=\sum_{i=1}^nR_iI(X_i>0).
    \label{eq2.2}
\end{equation}
here $R_{i}$ denotes the rank of $|X_i|$ among $\{|X_i|,i = 1,...,n\}$, where $|X_i|$ is the absolute value of $X_{i}$. Note that $E_{H_0}(U)=n(n+1)/4$, $Var_{H_0}(U)=n(n+1)(2n+1)/24$, and $(U-E_{H_0}(U))/\sqrt{Var(U_{H_0})}\overset{d}{\rightarrow}N(0,1)$. To handle the high dimensional problem, \cite{ouyang2022rank} proposed a sum-type statistic based on Wilcoxon signed-rank test, defined by:
\begin{equation}\label{eq3}
    M_n=p^{-1}\sum_{i=1}^{p}\{U_{i}-E_{H_0}(U_i) \}^2.
\end{equation}
where $U_i$ is the Wilcoxon signed-rank statistic from the $i$-th component of $p$ dimensional vector $\bm{X}$. Define $M_n^i=\{U_{i}-E_{H_0}(U_i) \}^2$, $i\in\{1,...,p\}$, it can be derived that $E_{H_0}(M_n^i)=n(n+1)(2n+1)/24$ and $Var_{H_0}(M_n^i)=(6n+5n^2-30n^3-25n^4+24n^5+20n^6)/1440$, then define the $R_n^i=(M_n^i-E_{H_0}(M_n^i))/\sqrt{Var_{H_0}(M_n^i)}$ . They proved the asymptotic normality
of $M_n$ under the following strong mixing and stationary assumptions.

\begin{assumption}\label{as1}
Assume that the strong stationary sequence $\{R_n^i\}$
satisfies the strong mixing condition. Let $\alpha(r)=sup\{\mathcal{F}_k^1,\mathcal{F}_{k+r}^p\}$, where $\mathcal{F}_m^n=\sigma\{m\leq R_n^i\leq n\}$ and $\alpha(\mathcal{F},\mathcal{G})=sup\{ |P(A\cap B)-P(A)P(B)|\}$.  The random sequence $R_n^i$
 is said to be "strongly mixing", if $\alpha(r)\rightarrow 0$ as $r\rightarrow \infty.$
\end{assumption}

\begin{assumption}\label{as2}
Suppose that $\sum_{r=1}^{\infty}\alpha(r)^{\delta/(2+\delta)}<\infty$ for some $\delta>0$, and for any $k\geq 0$, $\mathop{\lim}_{p\rightarrow \infty}\sum_{j=1}^{p-k}cov(R_n^j,R_n^{j+k})/(p+k)=\gamma(k)$ exists.
\end{assumption}

\cite{tony2014two} mentioned that sum-of-squares type statistics  does not perform as well in sparse cases and showed that max-type statistic performs pretty well under sparse alternatives.  Now we formally introduce our newly proposed max-type statistic, defined by:
\begin{equation}
    V_n=\max_{1\leq i\leq p}|V_n^i|.
\end{equation}
where $V_n^i\overset{def}{=}(U_i-E_{H_0}(U_i))/\sqrt{Var_{H_0}(U_i)}$. To derive the asymptotic distribution of $V_n$, additional assumptions need to be imposed. Thus we assume the following condition:
\begin{assumption}\label{as3}
    Let $\sigma_{ij}^n=cor(V_n^i,V_n^j)$, then $\mathop{lim}_{n\rightarrow\infty}\sigma_{ij}^n=\sigma_{ij}$, define $\mathbf{\Sigma}=(\sigma_{ij})_{1\leq i,j\leq p}$. For some $\rho\in(0,1)$, assume $\sigma_{ij}\leq \rho$ for all $1\leq i,j\leq p$. Suppose $\{\delta_p;p\geq1\}$ and $\{\zeta_p;p\geq1\}$ are positive constants with $\delta_p=o(1/\log p)$ and $\zeta=\zeta_p\rightarrow 0$ as $p\rightarrow \infty$. For $1\leq i\leq p$, define $B_{p,i}=\{1\leq j\leq p;|\sigma_{ij}|\geq\delta_p\}$ and $C_{p}=\{1\leq i\leq p;|B_{p,i}|\geq p^{\zeta}\}$. Assuming that $|C_p|/p$ as $p\rightarrow \infty$.
\end{assumption}
Note that Assumption \ref{as3} is the same as assumption (A1) in \cite{chen2022rank}.

\begin{thm}\label{th1}
Under the null hypothesis in (\ref{eq1}), we have following result:
\begin{itemize}
\item[(i)] Under assumptions \ref{as1}, \ref{as2} satisfied, $T_{sum}^{(1)}=\frac{\sqrt{p}(M_n-E_{H_0}(M_n^i))}{\sqrt{Var_{H_0}(M_n^i)\tau_1^2}}\rightarrow N(0,1)$, where $\tau_1^2=1+2\mathop{\sum}_{k=1}^{\infty}\gamma_1(k)$.
\item[(ii)] Under assumption \ref{as3}, let $T_{max}^{(1)}=V_n^2-2\log p+\log\log p$, we have:
\begin{equation}
    |P(T_{max}^{(1)}\leq y)-\exp\{-\pi^{-\frac{1}{2}}\exp\{ -y/2\} \} |=o(1).
\end{equation}
\item[(iii)] Under assumptions \ref{as1}-\ref{as3}, then $T_{sum}^{(1)}$ and $T_{max}^{(1)}$ are asymptotically independent.
\end{itemize}
\end{thm}

Theorem \ref{th1} (i) is from the Theorem 1 in \cite{ouyang2022rank}, where assumption \ref{as1}, \ref{as2} depict the dependent structure among the  rank-statistics of vector $\bm{X}$. When performing a $\alpha$ level hypothesis test based on $T_{sum}^{(1)}$, it rejects the null hypothesis for $1-\Phi^{-1}(T_{sum}^{(1)})\leq \alpha$, where $\Phi(x)$ denotes the c.d.f of standard normal distribution. And for the
 $T_{max}^{(1)}$ in Theorem \ref{th1} (ii), it rejects the null hypothesis when $1-G^{-1}(T_{max}^{(1)})\leq \alpha$, where $G(x)=\exp\{-\pi^{-\frac{1}{2}}\exp\{ -x/2\} \}$ is the c.d.f of Gumbel distribution. The technical proof of Theorem \ref{th1} (ii) is presented in appendix.

 With the asymptotic independence between $T_{sum}^{(1)}$ and $T_{max}^{(1)}$ in Theorem \ref{th1} (iii), we combine the statistics  by using Cauchy combination method in \cite{liu2019cauchy} :
 \begin{equation}
     T_{com}^{(1)}=1-C[0.5\tan\{(0.5-p^{(1)}_M)\pi\}+0.5\tan\{(0.5-p^{(1)}_S)\pi\} ].
 \end{equation}
 where
 \begin{gather*}
     p^{(1)}_S=1-\Phi(T_{sum}^{(1)}), p^{(1)}_M=1-G(T_{max}^{(1)}).
 \end{gather*}
Here $C(x)$ is the c.d.f of Cauchy distribution.
Thus we can perform the $\alpha$ level hypothesis test based on $T_{com}^{(1)}$ by rejecting null hypothesis when $T_{com}^{(1)}\leq \alpha$.

Next we consider the power function of $T_{com}^{(1)}$.
 \cite{li2023} proved that Cauchy combination test
would be more powerful than minimal p-value combination test, where minimal p-value combination test rejects the null hypothesis when $min(p_S, p_M)\leq 1-\sqrt{1-\alpha}\approx \alpha/2$. Thus we have:
\begin{align}
    \beta(T_{com}^{(1)},\alpha)&=P(T_{com}^{(1)}\leq \alpha)     \nonumber                             \\
    &\geq P(min(p_S, p_M)\leq 1-\sqrt{1-\alpha})\approx P(min(p_S, p_M)\leq \alpha/2) \nonumber\\
   &\geq max(\beta(T_{sum}^{(1)},\alpha/2), \beta(T_{max}^{(1)},\alpha/2) ).  \label{eq12}
\end{align}
We consider the power function under sparse and dense cases. In dense cases, the power function of $T_{sum}^{(1)}$ converges to 1
when $n\mu_i^2=O(p^{\alpha})$ with $\alpha\geq -1/2$ by Theorem 2 in \cite{ouyang2022rank}. In sparse cases, with the similar argument
in \cite{feng2022asymptotic}, the power function of $T_{max}^{(1)}$
converges to 1 provided the signal condition $\max_{1\leq i\leq p}\mu_i\gtrsim \sqrt{\log p/n}$ with $\log p=o(n)$.
\begin{thm}\label{th2}
    The power function of $T_{max}^{(1)}$
converges to 1 provided the signal condition $\max_{1\leq i\leq p}\mu_i\gtrsim \sqrt{\log p/n}$ with $\log p=o(n)$.
\end{thm}
Since the difference between the $\beta(T_{sum}^{(1)}, \alpha/2)$ and $\beta(T_{sum}^{(1)}, \alpha)$ should be negligible when $\alpha$ is relatively small and the same for $\beta(T_{sum}^{(1)}, \alpha)$, we can conclude that the power function of $T_{com}^{(1)}$ converges to 1 in both sparse and dense conditions according to (\ref{eq12}).


\begin{subsection}{Two sample test}
For the two-sample testing problem, we suppose there are $n$ samples $\{\bm{X}_1,...,\bm{X}_n\}$ drawn from  distribution $F(\bm{X}-\bm{\mu}_1)$ and $m$ samples $\{\bm{Y}_1,...,\bm{Y}_m\}$  from  distribution $F(\bm{Y}-\bm{\mu}_2)$.  We focus on testing the hypothesis problem
\begin{equation}\label{eq13}
  H_0:\bm{\mu}_1=\bm{\mu}_2 \ \ versus\  \ H_1:\bm{\mu}_1\neq \bm{\mu}_2.
\end{equation}
In $p=1$, the Wilcoxon-Mann-Whitney(WMW) statistic is defined as:
\begin{equation}
    U^{xy}=\sum_{i=1}^n R_{i}^{xy}-n(n+1)/2.
\end{equation}
Here $R_{i}^{xy}$ denotes the the rank of $X_{i}$ in the union of the sample $\{X_{1},...,X_{n},Y_{1},...,Y_{m}\}$. And $E_{H_0}(U^{xy})=mn/2$. Similar to the statistic in (\ref{eq3}), \cite{ouyang2022rank} also proposed a sum-type statistic for two sample test, which is defined as:

\begin{equation}
    M_n^{xy}=p^{-1}\sum_{i=1}^p (U_i^{xy}-E_{H_0}(U_i^{xy}))^2.
\end{equation}
where $U_i^{xy}$ is the WMW statistic of the i-th component in two sample test. Define $\gamma_n^i=\{U_{i}^{xy}-E_{H_0}(U_i^{xy}) \}^2$, $i\in\{1,...,p\}$ and we have $E_{H_0}(\gamma_n^i)=mn(m+n+1)/12$ and $Var_{H_0}(\gamma_n^i)=\{mn(5(m+n)+8)-3(m+n)(m+n+1)\}(m+n+1)mn/360$, then define the $\rho_n^i=(\gamma_n^i-E_{H_0}(\gamma_n^i))/\sqrt{Var_{H_0}(\gamma_n^i)}$ . With similar argument above, asymptotic normality
of $M_n^{xy}$ is proved.

Then we introduce our max-type statistic for two sample problem, defined as:
\begin{equation}
    V_n^{xy}=\max_{1\leq i\leq p}|\nu_n^i|.
\end{equation}
where $\nu_n^i\overset{def}{=}(U_i^{xy}-E_{H_0}(U_i^{xy}))/\sqrt{Var_{H_0}(U_i^{xy})}=
(U_i^{xy}-mn/2)/\sqrt{mn(m+n+1)/12}$. With the assumption \ref{as3} satisfied, we can derive the
asymptotic distribution of $V_n^{xy}$. Thus we have the following theorem:
\begin{thm}\label{th3}
    Under the null hypothesis in (\ref{eq13}), we have following result:
\begin{itemize}
\item[(i)] When assumptions \ref{as1}, \ref{as2} are satisfied with $R_n^i$ replaced by $\rho_n^i$, $T_{sum}^{(2)}=\frac{\sqrt{p}(M_n^{xy}-E_{H_0}(\gamma_n^i))}{\sqrt{Var_{H_0}(\gamma_n^i)\tau_2^2}}\rightarrow N(0,1)$, where $\tau_2^2=1+2\mathop{\sum}_{k=1}^{\infty}\gamma_2(k)$.
\item[(ii)] With assumption \ref{as3} and $V_n^i$ replaced by $\nu_n^i$, let $T_{max}^{(2)}=(V_n^{xy})^2-2\log p+\log\log p$, we have:
\begin{equation}
    |P(T_{max}^{(2)}\leq y)-\exp\{-\pi^{-\frac{1}{2}}\exp\{ -y/2\} \} |=o(1).
\end{equation}
\item[(iii)] With assumption \ref{as1} and $R_n^i$ replaced by $\rho_n^i$, then $T_{sum}^{(2)}$ and $T_{max}^{(2)}$ are asymptotically independent.
\end{itemize}
\end{thm}

Theorem \ref{th2} (i) is from the Theorem 3 in \cite{ouyang2022rank}. $T_{sum}^{(2)}$ performs a $\alpha$ level hypothesis test by rejecting the null hypothesis for $1-\Phi^{-1}(T_{sum}^{(2)})\leq \alpha$. And for the
 $T_{max}^{(2)}$ in Theorem \ref{th3} (ii), it performs a $\alpha$ level hypothesis test by rejecting the null hypothesis for $1-G^{-1}(T_{max}^{(2)})\leq \alpha$. The technical proof of Theorem \ref{th3} (ii)
 is presented in the appendix.

 Since Theorem \ref{th3} (iii) provides the asymptotic independence between $T_{sum}^{(2)}$ and $T_{max}^{(2)}$, we propose a Cauchy combination test $T_{com}^{(2)}$ for two sample test, defined as:
 \begin{equation}
     T_{com}^{(2)}=1-C[0.5\tan\{(0.5-p^{(2)}_M)\pi\}+0.5\tan\{(0.5-p^{(2)}_S)\pi\} ].
 \end{equation}

 where
 \begin{gather*}
     p_S^{(2)}=1-\Phi(T_{sum}^{(2)}),p_M^{(2)}=1-G(T_{max}^{(2)}).
 \end{gather*}

Thus we perform the $\alpha$ level hypothesis test based on $T_{com}^{(2)}$ by rejecting null hypothesis when $T_{com}^{(2)}\leq \alpha$.

Then we consider the power function of $T_{com}^{(2)}$. With the discussion in (\ref{eq12}), we consider $max(\beta(T_{sum}^{(2)},\alpha/2), \beta(T_{max}^{(2)},\alpha/2) )$
directly. In dense cases, by the Theorem 4 in \cite{ouyang2022rank}, power function of $T_{sum}^{(2)}$ converges to 1 when $n(\mu_{1i}-\mu_{2i})^2=O(p^{\alpha})$ with $\alpha\geq -1/2$, here $\mu_{ki}$ denotes the ith coordinate of $\mu_k$, $k$=1,2. In sparse cases, the power function of $T_{max}^{(2)}$
converges to 1 provided the signal condition $\max_{1\leq i\leq p}(\mu_{1i}-\mu_{2i})\gtrsim \sqrt{(m+n)\log p/mn}$ with $\log p=o(mn/(m+n))$.
\begin{thm}\label{th4}
    The power function of $T_{max}^{(2)}$
converges to 1 provided the signal condition $\max_{1\leq i\leq p}(\mu_{1i}-\mu_{2i})\gtrsim \sqrt{(m+n)\log p/mn}$ with $\log p=o(mn/(m+n))$.
\end{thm}
Thus we can conclude that the power function of $T_{com}^{(2)}$ converges to 1 in either sparse or dense conditions.

\end{subsection}

\section{Simulation}
In this section, we carried out series of simulations to evaluate the performance of our proposed method. We incorporated various methods into our study:
\begin{itemize}
\item our max-type method $T_{max}^{(1)}$ in one sample, $T_{max}^{(2)}$ in two sample, referred as
MAX1;
\item sum-type method $T_{sum}^{(1)}$ for one sample, $T_{sum}^{(2)}$ for two sample in \cite{ouyang2022rank}, referred as
SUM1;
\item our combination test $T_{com}^{(1)}$ for one sample, $T_{com}^{(2)}$ for two sample, referred as
COM1;
\item max-type method proposed by \cite{feng2022asymptotic}
, referred as
MAX2;
\item sum-type method proposed by \cite{srivastava2008test}, referred as
SUM2;
\item combination test proposed by \cite{feng2022asymptotic}, referred as
COM2.
\end{itemize}
\label{sec:verify}
\begin{subsection}{one sample test}
\end{subsection}

Following the setting in \cite{ouyang2022rank} and \cite{feng2022asymptotic}, we generate
$n$ samples $\{\bm{x}_1,...,\bm{x}_n\}$ with mean vector $\bm{\mu}=(\mu_1,...,\mu_p)$ and covariance matrix $\mathbf{\Sigma}$. We consider data from two distributions: multivariate normal distribution and the multivariate $t$ distribution $t_3$. We generate data with dimension $n=100,200$ and sample sizes $p=200,400,600$. In order to compare the performance of above methods under different dependent structures, we consider the following three scenarios:

(1)Independent structure: $\mathbf{\Sigma}=I_{(p\times p)}$.

(2)AR(1) model with weak correlation: $\mathbf{\Sigma}=(0.3^{|i-j|})_{1\leq i,j\leq p}$.

(3)AR(1) model with strong correlation: $\mathbf{\Sigma}=(0.6^{|i-j|})_{1\leq i,j\leq p}$.

We first check the performance of empirical sizes under null cases,i.e. $\bm{\mu}=0$. Table \ref{tab:my_label}, Table \ref{tab:my_label2}, table \ref{tab:my_label3} display the empirical size
under scenario (1), scenario (2), scenario (3), respectively. Each result is obtained by 1000 simulations. The simulated results show that MAX1, SUM1, COM1 perform pretty well under diverse scenarios, while MAX2, SUM2, COM2 behave quite conservative with $t_3$ distribution.

\begin{figure}
    \centering
    \includegraphics[width=\textwidth]{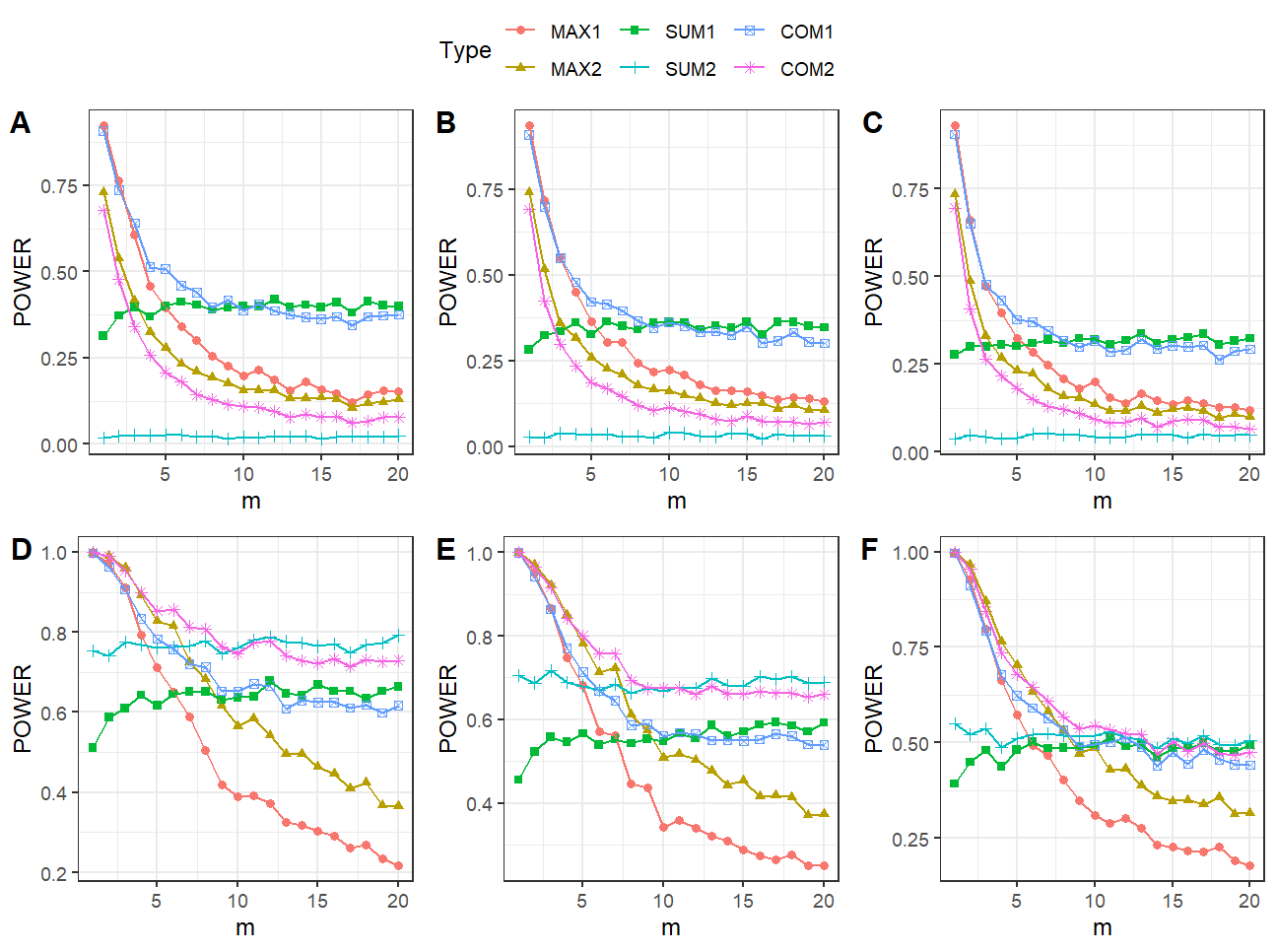}
 \caption{POWER curves of MAX1,SUM1,COM1,MAX2,SUM2,COM2 as the number of nonzero component $m$ increases under three scenarios with two distribution in one sample test. A, B, C for $t_3$ distribution under scenario (1), scenario (2), scrnario (3), respectively. D, E, F for normal distribution under scenario (1), scenario (2), scrnario (3), respectively.}
    \label{fig1}
\end{figure}

For power comparison, we follow the setting in \cite{feng2022asymptotic} and set $n$=100, $p$=200. For the alternative cases, $\bm{\mu}$ is set to $(0,0,0...,\sqrt{0.5/m},...,\sqrt{0.5/m})$, where $m$ denotes the number of nonzero components.  Figure \ref{fig1} displays the POWER curves of the above methods as the number of nonzero component $m$ increases under three scenarios with multivariate normal distribution and multivariate $t_3$ distribution. For the multivariate $t_3$ distribution, the power of both MAX1 and MAX2 methods decreases as $m$ increases. This is expected as max-type tests are more effective in sparse cases. On the other hand, the power of SUM1 and SUM2 methods remains stable, as we have fixed the signal $\bm{\mu}^T\mathbf{\Sigma}\bm{\mu}$.
Our findings indicate that both MAX1 and SUM1 tests are more powerful than their counterparts, MAX2 and SUM2, respectively. As a result, COM1 outperforms COM2 in all cases. This is not surprising as rank-based tests have been shown to outperform the classic $t-$test for heavy-tailed distributions.
The COM1 test exhibits superior performance in nearly all tests. When the signal is very sparse, i.e., $m<5$, COM1 performs similarly to MAX1. When the signal is very dense, i.e., $m>10$, COM1 performs similarly to SUM1. However, for $5\le m\le 10$, COM1 outperforms both MAX1 and SUM1.
These results align with theoretical predictions and highlight the superiority of the COM1 method.
It is worth mentioning that under normal distribution, COM1 did not perform as well as COM2 while the difference between COM1 and COM2 is not significant, which is natural since COM2 is proposed based on the assumption of normality.


\begin{table}[t]
    \caption{Sizes of one sample tests with Scenario (1)}
    \centering
    \begin{tabular}{c|c|c}
    \hline
    \hline
\qquad   Distribution & Normal&$t_3$ \\
        \hline
\qquad   p & 200\ \   \ 400\ \   \ 600&200\ \ \ 400\ \   \ 600\\
      \hline
n=100\ \ \      MAX1\qquad\  & 0.036\ 0.033\ 0.033   &0.030\ 0.031\ 0.040\\

 \qquad    SUM1&0.055\ 0.060\ 0.067&0.046\ 0.068\ 0.057 \\

  \qquad    COM1&0.059\ 0.055\ 0.066&0.043\ 0.065\ 0.059 \\

 \qquad    MAX2&0.082\ 0.065\ 0.088&0.022\ 0.035\ 0.036\\

 \qquad    SUM2&0.063\ 0.071\ 0.076&0.001\ 0.000\ 0.000\\

 \qquad    COM2&0.067\ 0.063\ 0.085&0.013\ 0.012\ 0.015\\
     \hline
n=200\ \ \      MAX1\qquad\  &0.043\ 0.035\ 0.035& 0.031\ 0.042\ 0.040\\

 \qquad     SUM1&0.061\ 0.046\ 0.044&0.053\ 0.049\ 0.065\\

 \qquad     COM1&0.066\ 0.044\ 0.041& 0.069\ 0.065\ 0.059\\

 \qquad   MAX2&0.060\ 0.055\ 0.065& 0.035\ 0.033\ 0.036\\

 \qquad     SUM2&0.058\ 0.052\ 0.055&0.004\ 0.000\ 0.000\\

 \qquad     COM2&0.062\ 0.055\ 0.058& 0.014\ 0.015\ 0.014\\
  \hline
  \hline
    \end{tabular}
    \label{tab:my_label}
\end{table}

\begin{table}[t]
\caption{Sizes of one sample tests with Scenario (2)}
    \centering
    \begin{tabular}{c|c|c}
    \hline
    \hline
\qquad        distribution & Normal&$t_3$ \\
        \hline
 \qquad     p & 200\ 400\ 600&200\ 400\ 600\\
      \hline
n=100\ \ \      MAX1\qquad\  & 0.038\ 0.034\ 0.033   &0.027\ 0.023\ 0.029\\

\qquad      SUM1&0.044\ 0.041\ 0.040&0.041\ 0.030\ 0.045 \\

\qquad      COM1&0.053\ 0.046\ 0.040&0.039\ 0.036\ 0.039 \\

\qquad      MAX2&0.070\ 0.074\ 0.086&0.026\ 0.033\ 0.041\\

\qquad      SUM2&0.060\ 0.059\ 0.061&0.005\ 0.000\ 0.000\\

\qquad      COM2&0.064\ 0.066\ 0.077&0.012\ 0.013\ 0.020\\
    \hline
     n=200\ \ \      MAX1\qquad\  &0.037\ 0.037\ 0.037& 0.039\ 0.039\ 0.032\\

 \qquad     SUM1&0.042\ 0.041\ 0.053&0.052\ 0.041\ 0.035\\

 \qquad     COM1&0.057\ 0.053\ 0.059& 0.069\ 0.055\ 0.042\\

 \qquad   MAX2&0.054\ 0.055\ 0.062& 0.034\ 0.023\ 0.031\\

\qquad      SUM2&0.056\ 0.066\ 0.063&0.004\ 0.000\ 0.000\\

\qquad      COM2&0.048\ 0.057\ 0.055& 0.019\ 0.007\ 0.016
  \\
  \hline
  \hline

    \end{tabular}
    \label{tab:my_label2}
\end{table}

\begin{table}[t]
\caption{Sizes of one sample tests with Scenario(3)}
    \centering
    \begin{tabular}{c|c|c}
    \hline
    \hline
    \qquad    distribution & Normal&$t_3$ \\
        \hline
 \qquad     p & 200\ 400\ 600&200\ 400\ 600\\
      \hline
n=100\ \ \      MAX1\qquad\  & 0.043\ 0.030\ 0.036   &0.032\ 0.036\ 0.032\\

 \qquad     SUM1&0.044\ 0.064\ 0.074&0.065\ 0.062\ 0.044 \\

 \qquad     COM1&0.070\ 0.064\ 0.068&0.070\ 0.064\ 0.049 \\

 \qquad     MAX2&0.069\ 0.066\ 0.083&0.031\ 0.031\ 0.041\\

 \qquad     SUM2&0.064\ 0.062\ 0.064&0.006\ 0.000\ 0.000\\

 \qquad     COM2&0.070\ 0.068\ 0.079&0.022\ 0.019\ 0.017\\
    \hline
 n=200\ \ \      MAX1\qquad\  &0.041\ 0.030\ 0.038& 0.035\ 0.033\ 0.046\\

 \qquad     SUM1&0.074\ 0.065\ 0.068&0.055\ 0.057\ 0.079\\

 \qquad     COM1&0.074\ 0.068\ 0.072& 0.068\ 0.069\ 0.074\\

 \qquad   MAX2&0.055\ 0.043\ 0.057& 0.031\ 0.030\ 0.029\\

 \qquad     SUM2&0.056\ 0.059\ 0.059&0.009\ 0.003\ 0.001\\

 \qquad     COM2&0.060\ 0.053\ 0.054& 0.017\ 0.018\ 0.015
  \\
  \hline
  \hline

    \end{tabular}

    \label{tab:my_label3}
\end{table}

\subsection{Two sample test}
    In this part we present the numerical results of two sample test. Similar to the setting in one sample test, we generate
$n$ samples $\{\bm{x}_1,...,\bm{x}_n\}$ with mean vector $\bm{\mu}_1=(\mu_{11},...,\mu_{1p})$,  $n$ samples $\{\bm y_1,...,\bm y_n\}$ with mean vector $\bm{\mu}_2=(\mu_{21},...,\mu_{2p})$ with the same covariance matrix $\mathbf{\Sigma}$ as $\bm x_i$. We consider data from two distributions:  multivariate normal distribution and multivariate  $t$ distribution $t_3$. We generate data with dimension $p=200,400,600$ and sample sizes $n=100,200$ . Here we only present the results of scenario (2) to avoid tautology, i.e. $\mathbf{\Sigma}=(0.3^{|i-j|})_{1\leq i,j\leq p}$. The performance of each test under the other two structure of covariance matrix are similar. So we omit it here.

We display the performance of empirical sizes of each test in Table \ref{tab:my_label4}.  The simulated results indicate that MAX1, SUM1, COM1, MAX2, SUM2, COM2 perform pretty well under both multivariate normal distribution. However, for the multivariate $t_3$-distribution, the sizes of the mean-based test procedures--MAX2,SUM2,COM2 are lower than the nominal level. While, our proposed test procedures MAX2 and COM2 can control the empirical sizes in all cases, which show the robustness of our methods.

\begin{figure}[t]
    \centering
    \includegraphics[width=0.8\textwidth,height=0.4\textwidth]{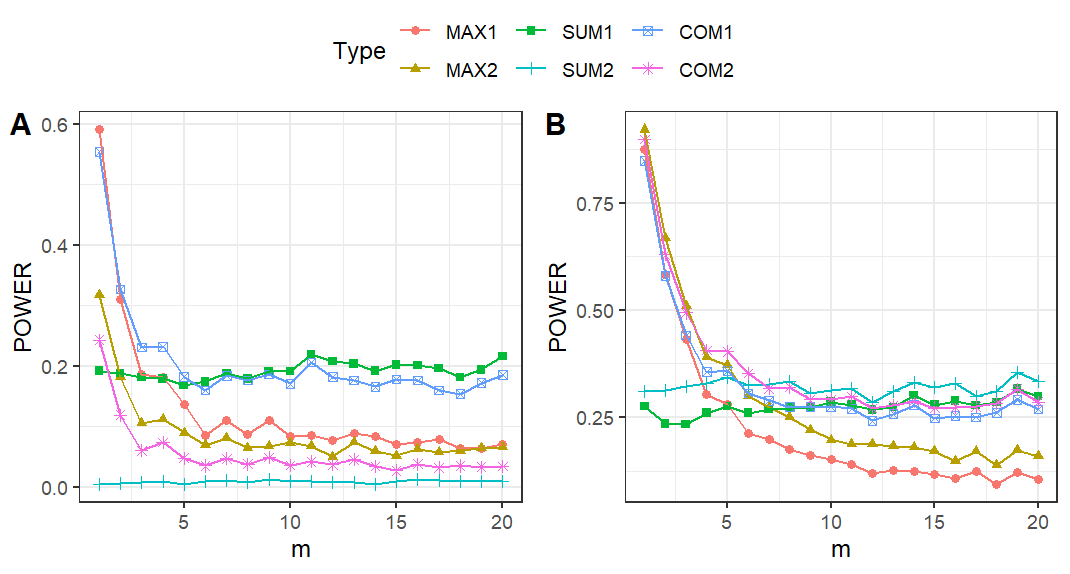}
    \caption{POWER curves of MAX1,SUM1,COM1,MAX2,SUM2,COM2 as the number of
nonzero component $m$ increases under scenario (2) with two distributions in two sample test. A for $t_3$ distribution, B for normal distribution.}
    \label{fig2}
\end{figure}

For the two sample test, we let $\bm{\mu}_1=(0,0,0...,\sqrt{0.5/m},...,\sqrt{0.5/m})$ and set $\bm{\mu}_2=0$, where $m$ denotes the number of nonzero components . Here we only present the simulation results of scenario (2) with multivariate normal distribution and $t_3$ distribution. Figure \ref{fig2} displays the POWER curves of the above methods as the number of nonzero component $m$ increases under  scenario (2). It can be seen that under normal conditions, the difference between COM1 and COM2 is not significant, while for the multivariate $t_3$-distribution, COM1 outperforms COM2 significantly. Hence we conclude that COM1 maintains the robustness from COM2 proposed by \cite{feng2022asymptotic} and outperforms COM2 substantially.


\begin{table}[t]
\caption{Sizes of two sample tests with Scenario (2)}
    \centering
    \begin{tabular}{c|c|c}
    \hline
    \hline
    \qquad    distribution & Normal&$t_3$ \\
        \hline
 \qquad     $p$ & 200\ 400\ 600&200\ 400\ 600\\
      \hline
$n=100$\ \ \      MAX1\qquad\  & 0.032\ 0.033\ 0.045   &0.037\ 0.037\ 0.021\\

 \qquad     SUM1&0.063\ 0.450\ 0.055&0.061\ 0.063\ 0.048 \\

 \qquad     COM1&0.069\ 0.050\ 0.063&0.049\ 0.054\ 0.042 \\

 \qquad     MAX2&0.050\ 0.055\ 0.071&0.028\ 0.042\ 0.024\\

 \qquad     SUM2&0.059\ 0.046\ 0.056&0.001\ 0.000\ 0.000\\

 \qquad     COM2&0.044\ 0.051\ 0.072&0.013\ 0.023\ 0.010\\
    \hline
 $n=200$\ \ \      MAX1\qquad\  &0.049\ 0.035\ 0.050& 0.038\ 0.034\ 0.050\\

 \qquad     SUM1&0.062\ 0.058\ 0.071&0.060\ 0.060\ 0.069\\

 \qquad     COM1&0.076\ 0.053\ 0.067& 0.054\ 0.048\ 0.074\\

 \qquad   MAX2&0.060\ 0.053\ 0.065\ & 0.028\ 0.030\ 0.038\\

 \qquad     SUM2&0.054\ 0.048\ 0.056\ &0.001\ 0.001\ 0.001\\

 \qquad     COM2&0.057\ 0.052\ 0.055\ & 0.010\ 0.012\ 0.014
  \\
  \hline
  \hline

    \end{tabular}

    \label{tab:my_label4}
\end{table}

\section{Conclusion}
\label{sec:conc}

In this study, we introduce a rank-based testing method that exhibits robustness to both signal sparsity and heavy-tailed distributions. Our theoretical findings and simulation experiments collectively highlight the advantages of the methods we propose. Moreover, \cite{he2021asymptotically} suggested a set of U-statistics as an unbiased estimator for the $L_q$-norms of the mean in high-dimensional mean testing problems, demonstrating that U-statistics of varying finite orders are asymptotically independent and normally distributed. The question of how to develop a rank-based U-statistic family for high-dimensional mean testing problems is intriguing and warrants further exploration.

\section*{Appendix}
We present the detailed proof of Theorem \ref{th1} (ii), (iii) and Theorem \ref{th2}, where Theorem \ref{th1} (ii) adopts a similar method to \cite{chen2022rank}. And the proof of Theorem \ref{th3} (ii), (iii) and Theorem \ref{th4} can be derived similarly.
\subsection*{Appendix A: Proof of Theorem \ref{th1}}
\subsubsection*{Proof of Theorem \ref{th1} (ii)}
Define $z=(2\log p-\log\log p+y)^{\frac{1}{2}}$. Under the  null cases, we can rewrite the expression in \ref{eq2.2} as:
\begin{equation*}
    U=\sum_{i=1}^nR_iI(X_i>0)=\sum_{j=1}^n jW_j
\end{equation*}
here $W_j=I(|X_{j}| \ corresponds\  to\  some\  positive\  X_{j'} )$. Then by the above representation of $U$ and Hajek-Sidak’s CLT, we have
\begin{equation*}
    P(|V_i|>z)=2\{1-\Phi(z)\}(1+o(1))\sim \frac{1}{\sqrt{\pi}}\frac{e^{-\frac{y}{2}}}{p}.
\end{equation*}

Thus,
\begin{equation*}
    p(\mathop{\max}_{i\in C_p} |V_i|>z)\leq |C_p|P(|V_i|>z)\rightarrow 0.
\end{equation*}

 Set $D_{p}=\{1\leq i\leq p;|B_{p,i}|\leq p^{\zeta}\}$. By assumption, $|D_p|/p\rightarrow 1\  as\  p\rightarrow \infty.$ Since
 \begin{equation*}
     P(\mathop{\max}_{i\in D_p}|V_i|>z)\leq P(\mathop{\max}_{1\leq i\leq p}|V_i|>z)\leq P(\mathop{\max}_{i\in D_p}|V_i|>z)+P(\mathop{\max}_{i\in C_p}|V_i|>z),
 \end{equation*}
it's enough to show
\begin{equation*}
  \mathop{\lim}_{p\rightarrow\infty}   P(\mathop{\max}_{i\in D_p}|V_i|>z)=1-\exp(-\pi^{-\frac{1}{2}}\exp\{ -y/2\} ).
\end{equation*}
Here define:
\begin{equation*}
    \alpha_t=\sum^{*}P(|V_{i1}|>z,...,|V_{it}|>z).
\end{equation*}
for $1\leq t\leq p$, where the sum runs over all $i_1\leq ...\leq i_t$, and $i_1,...,i_t\in D_p$. First we prove
\begin{equation}
    \mathop{\lim}_{p\rightarrow \infty}\alpha_t=\frac{1}{t!}\pi^{-\frac{t}{2}}e^{-\frac{ty}{2}}
    \label{eq4.7}
\end{equation}
for each $t\geq 1$.

By the representation of $V_i$, for $1\leq j\leq n$, $W_j$ is bounded by a constant when $n$ is sufficiently large. Thus we can apply the Theorem1.1 in Zaitsev (1987) and get the following inequalities:

\begin{gather*}
    \sum^*P(|Z_{i1}|>z+\epsilon_n/\log p,...,|Z_{it}|>z+\epsilon_n/\log p)-(\begin{array}{c}
        |D_p| \\
         p
    \end{array})
    c_1t^{5/2}\exp(\frac{-n^{1/2}\epsilon_n}{c_2t^3(\log p)})\\
    \leq \sum^* P(|V_{i1}|>z,...,|V_{it}|>z) \\
    \leq \sum^*P(|Z_{i1}|>z+\epsilon_n/\log p,...,|Z_{it}|>z+\epsilon_n/\log p)+(\begin{array}{c}
        |D_p| \\
         p
    \end{array})
    c_1t^{5/2}\exp(\frac{-n^{1/2}\epsilon_n}{c_2t^3(\log p)})
\end{gather*}
Here $(Z_{i1},...,Z_{it}
)$  has the same
covariance matrix with $(V_{i1},...,V_{it}
)$ and follows normal distribution with zero mean. By the proof of Theorem 2 in Feng et al. (2022a), we
have
\begin{gather*}
\sum^*P(|Z_{i1}|>z+\epsilon_n/\log p,...,|Z_{it}|>z+\epsilon_n/\log p)\rightarrow \frac{1}{t!}\pi^{-\frac{t}{2}}e^{-\frac{ty}{2}}\\
\sum^*P(|Z_{i1}|>z-\epsilon_n/\log p,...,|Z_{it}|>z-\epsilon_n/\log p)\rightarrow \frac{1}{t!}\pi^{-\frac{t}{2}}e^{-\frac{ty}{2}}
\end{gather*}
with $\epsilon_n\rightarrow 0$ and $p \rightarrow \infty$. While we have:
\begin{equation*}
(\begin{array}{c}
        |D_p| \\
        t
    \end{array})
    c_1t^{5/2}\exp(\frac{-n^{1/2}\epsilon_n}{c_2t^3(\log p)^{1/2}})\leq C(\begin{array}{c}
       p \\
       t
    \end{array})t^{5/2}\exp(\frac{-n^{1/2}\epsilon_n}{c_2t^3(\log p)^{1/2}})
\end{equation*}
for $\epsilon_n\rightarrow 0$  slowly enough . Hence we get (\ref{eq4.7}) proved.

Then by Bonferroni inequality,
\begin{equation}
    \sum_{t=1}^{2k}(-1)^{t-1}\alpha_t\leq P(\mathop{\max}_{i\in D_p}|V_i|>z)\leq\sum_{t=1}^{2k+1}(-1)^{t-1}\alpha_t.
\end{equation}
Let $k\rightarrow\infty$ and $p\rightarrow \infty$ , using the Taylor expansion making the proof completed.

\subsubsection*{Proof of Theorem \ref{th1} (iii)}
When assumption \ref{as3} holds, all the conditions required for Theorem 1.1 in \cite{hsing1995note} are met. Thus the asymptotic independence between the $T_{max}^{(1)}$ and $T_{sum}^{(1)}$ is derived.

\subsection*{Appendix B: Proof of Theorem \ref{th2}}

\cite{hettmansperger1984statistical} presented the moment properties of $U_i$ under local alternatives in Theorem 2.5.1:
\begin{align*}
    E_{H_1}(U_i)&=\frac{n(n-1)p_{2i}}{2}+np_{1i} \\
   Var_{H_1}(U_i)&=np_{1i}(1-p_{1i})+\frac{n(n-1)p_{2i}(1-p_{2i})}{2}+2n(n-1)(p_{3i}-p_{1i}p_{2i})+n(n-1)(n-2)(p_{4i}-p_{2i}^2),
   \end{align*}
    where
 \begin{align*}
 p_{1i}=F_i(\mu_i),\ p_{2i}=\int F_i(2\mu_i+x)f_i(x)dx,
    p_{3i}=(p_{1i}^2+p_{2i})/2,
    p_{4i}=\int \{F_i(2\mu_i+x)\}^2f_i(x)dx,
\end{align*}
and $F_i(x), f_i(x)$ are the marginal c.d.f and p.d.f of $i-$th component of $\bm{X}$. Thus, if $\mu_i=C\sqrt{\log p/n}$, we have $E_{H_1}(U_i)=\frac{n(n+1)}{4}+n(n-1)\mu_i \int f^2_i(x)dx+o(n^2\mu_i)$ and $Var_{H_1}(U_i)=Var_{H_0}(U_i)(1+o(1))$.

Taking the same procedure as the proof of Theorem \ref{th1}(ii), we have
\begin{align*}
 P\left(\max_{1\leq i\leq p}\left(\frac{U_i-E_{H_1}(U_i)}{\sqrt{Var_{H_1}(U_i)}}\right)^2-(2\log p-\log\log p)\le x\right)\to G(x)
\end{align*}
Hence we have:
\begin{align*}
 P\left(\max_{1\leq i\leq p}\left(\frac{U_i-E_{H_1}(U_i)}{\sqrt{Var_{H_1}(U_i)}}\right)^2\leq(2\log p-1/2\log\log p)\right)\to 1,
\end{align*}
when taking $x=1/2\log\log p$.
By the triangle inequality, we have:
\begin{align*}
\max_{1\leq i\leq p}\left(\frac{U_i-E_{H_0}(U_i)}{\sqrt{Var_{H_0}(U_i)}}\right)^2\geq &
\max_{1\leq i\leq p}\left(\frac{E_{H_0}(U_i)-E_{H_1}(U_i)}{\sqrt{Var_{H_0}(U_i)}}\right)^2-\max_{1\leq i\leq p}\left(\frac{U_i-E_{H_1}(U_i)}{\sqrt{Var_{H_0}
(U_i)}}\right)^2 \nonumber\\
\geq& \max_{1\leq i\leq p}\left(\frac{n(n-1)\mu_i\int f_i^2(x)dx+o(n^2\mu_i)}{\sqrt{Var_{H_0}(U_i)}}\right)^2  -(2\log p-1/2\log\log p),
\end{align*}
with probability approaching 1. Hence $T_{max}^{(1)}$ is consistent provided  $\max_{1\leq i\leq p}\mu_i\gtrsim \sqrt{\log p/n}$.

\subsection*{Appendix C: Proof of Theorem \ref{th3}}
The proof of Theorem\ref{th2} (ii) can be similarly derived as for Theorem\ref{th1} (ii) with the asymptotic normality of $U_i^{xy}$ and thus the details are omitted here.
The proof of Theorem\ref{th2} (ii) can be similarly derived as for Theorem \ref{th2} (ii).

\subsection*{Appendix D: Proof of Theorem \ref{th4}}
\cite{hettmansperger1984statistical} presented the moment properties of $\nu_i$ under local alternatives in Theorem 3.5.1:
\begin{gather}
    E_{H_1}(\nu_i)=mnp_1 \nonumber \\
  Var_{H_1}(\nu_i)=mn(p_1-p_1^2)+mn(n-1)(p_2-p_1^2)+mn(m-1)(p_3-p_1^2),\nonumber\\
    where\ \ \ \ \ \qquad \qquad p_1=\int(1-F(x-\mu_2))f(x-\mu_1)dx,  \qquad\qquad\qquad\qquad ,\nonumber\\
    p_2=\int\{1-F(x-\mu_2)\}^2f(x-\mu_1)dx,\nonumber\\
    p_3=\int F(x-\mu_1)^2f(x-\mu_2)dx.
\end{gather}
With the similar discussion in the proof of Theorem \ref{th2}, $T_{max}^{(2)}$ is consistent provided that $\max_{1\leq i\leq p}(\mu_{1i}-\mu_{2i})\gtrsim \sqrt{(m+n)\log p/mn}$.

\bigskip

%


@article{ouyang2022rank,
  title={A rank-based high-dimensional test for equality of mean vectors},
  author={Ouyang, Yanyan and Liu, Jiamin and Tong, Tiejun and Xu, Wangli},
  journal={Computational Statistics \& Data Analysis},
  volume={173},
  pages={107495},
  year={2022},
  publisher={Elsevier}
}

@article{tony2014two,
  title={Two-sample test of high dimensional means under dependence},
  author={Cai, Tony and Liu, Weidong and Xia, Yin},
  journal={Journal of the Royal Statistical Society Series B: Statistical Methodology},
  volume={76},
  number={2},
  pages={349--372},
  year={2014},
  publisher={Oxford University Press}
}

@article{feng2022asymptotic,
  title={Asymptotic independence of the sum and maximum of dependent random variables with applications to high-dimensional tests},
  author={Feng, Long and Jiang, Tiefeng and Li, Xiaoyun and Liu, Binghui},
  journal={arXiv preprint arXiv:2205.01638},
  year={2022}
}

@article{fan2015power,
  title={Power enhancement in high-dimensional cross-sectional tests},
  author={Fan, Jianqing and Liao, Yuan and Yao, Jiawei},
  journal={Econometrica},
  volume={83},
  number={4},
  pages={1497--1541},
  year={2015},
  publisher={Wiley Online Library}
}

@article{xu2016adaptive,
  title={An adaptive two-sample test for high-dimensional means},
  author={Xu, Gongjun and Lin, Lifeng and Wei, Peng and Pan, Wei},
  journal={Biometrika},
  volume={103},
  number={3},
  pages={609--624},
  year={2016},
  publisher={Oxford University Press}
}

@article{li2023,
title = {The Cauchy Combination Test under Arbitrary Dependence Structures},
journal = {The American Statistician},
volume = {77},
number = {2},
pages = {134-142},
year = {2023},
doi = {10.1080/00031305.2022.2116109},
URL = {https://doi.org/10.1080/00031305.2022.2116109},
author = {Mingya Long and Zhengbang Li and Wei Zhang and Qizhai Li}
}

@article{he2021asymptotically,
  title={Asymptotically independent U-statistics in high-dimensional testing},
  author={He, Yinqiu and Xu, Gongjun and Wu, Chong and Pan, Wei},
  journal={The Annals of Statistics},
  volume={49},
  number={1},
  pages={154--181},
  year={2021},
  publisher={Institute of Mathematical Statistics}
}

@article{chen2022rank,
  title={Rank Based Tests for High Dimensional White Noise},
  author={Chen, Dachuan and Feng, Long},
  journal={arXiv preprint arXiv:2204.08402},
  year={2022}
}

@article{liu2019cauchy,
  title={Cauchy combination test: a powerful test with analytic p-value calculation under arbitrary dependency structures},
  author={Liu, Yaowu and Xie, Jun},
  journal={Journal of the American Statistical Association},
  year={2019},
  publisher={Taylor \& Francis}
}

@book{hettmansperger1984statistical,
  title={Statistical inference based on ranks},
  author={Hettmansperger, Thomas P},
  year={1984},
  publisher={John Wiley \& Sons}
}

@book{newbold2013statistics,
  title={Statistics for business and economics},
  author={Newbold, Paul and Carlson, William L and Thorne, Betty M},
  year={2013},
  publisher={Pearson}
}

@article{bai1996effect,
  title={Effect of high dimension: by an example of a two sample problem},
  author={Bai, Zhidong and Saranadasa, Hewa},
  journal={Statistica Sinica},
  pages={311--329},
  year={1996},
  publisher={JSTOR}
}

@article{srivastava2008test,
  title={A test for the mean vector with fewer observations than the dimension},
  author={Srivastava, Muni S and Du, Meng},
  journal={Journal of Multivariate Analysis},
  volume={99},
  number={3},
  pages={386--402},
  year={2008},
  publisher={Elsevier}
}

@article{chen2010two,
  title={A two-sample test for high-dimensional data with applications to gene-set testing},
  author={Chen, Song Xi and Qin, Ying-Li},
  year={2010}
}

@article{miller1991basic,
  title={Basic statistical methods for analytical chemistry. Part 2. Calibration and regression methods. A review},
  author={Miller, James N},
  journal={Analyst},
  volume={116},
  number={1},
  pages={3--14},
  year={1991},
  publisher={The Royal Society of Chemistry}
}

@article{cheung2009students,
  title={Students’ attitudes toward chemistry lessons: The interaction effect between grade level and gender},
  author={Cheung, Derek},
  journal={Research in Science Education},
  volume={39},
  pages={75--91},
  year={2009},
  publisher={Springer}
}

@article{feng2016multivariate,
  title={Multivariate-sign-based high-dimensional tests for the two-sample location problem},
  author={Feng, Long and Zou, Changliang and Wang, Zhaojun},
  journal={Journal of the American Statistical Association},
  volume={111},
  number={514},
  pages={721--735},
  year={2016},
  publisher={Taylor \& Francis}
}

@article{wang2015high,
  title={A high-dimensional nonparametric multivariate test for mean vector},
  author={Wang, Lan and Peng, Bo and Li, Runze},
  journal={Journal of the American Statistical Association},
  volume={110},
  number={512},
  pages={1658--1669},
  year={2015},
  publisher={Taylor \& Francis}
}

@article{hsing1995note,
  title={A note on the asymptotic independence of the sum and maximum of strongly mixing stationary random variables},
  author={Hsing, Tailen},
  journal={The Annals of Probability},
  pages={938--947},
  year={1995},
  publisher={JSTOR}
}

@article{chen2019two,
  title={Two-sample and ANOVA tests for high dimensional means},
  author={Chen, Song Xi and Li, Jun and Zhong, Ping-Shou},
  journal={The Annals of Statistics},
  volume={47},
  number={3},
  pages={1443--1474},
  year={2019},
  publisher={JSTOR}
}

@article{park2013test,
  title={A test for the mean vector in large dimension and small samples},
  author={Park, Junyong and Ayyala, Deepak Nag},
  journal={Journal of Statistical Planning and Inference},
  volume={143},
  number={5},
  pages={929--943},
  year={2013},
  publisher={Elsevier}
}

@article{gregory2015two,
  title={A two-sample test for equality of means in high dimension},
  author={Gregory, Karl Bruce and Carroll, Raymond J and Baladandayuthapani, Veerabhadran and Lahiri, Soumendra N},
  journal={Journal of the American Statistical Association},
  volume={110},
  number={510},
  pages={837--849},
  year={2015},
  publisher={Taylor \& Francis}
}

@article{feng2015two,
  title={Two-sample Behrens-Fisher problem for high-dimensional data},
  author={Feng, Long and Zou, Changliang and Wang, Zhaojun and Zhu, Lixing},
  journal={Statistica Sinica},
  pages={1297--1312},
  year={2015},
  publisher={JSTOR}
}
\end{document}